# ANISOTROPY OF DC ELECTRIC FIELD INFLUENCE ON ACOUSTIC WAVE PROPAGATION IN PIEZOELECTRIC PLATE

S.I. Burkov[1], O.P. Zolotova[1], B.P. Sorokin[2], K.S. Aleksandrov[3]

*1) Siberian Federal University, 79 Svobodny ave., Krasnoyarsk, 660041, Russia*
*2) Technological Institute for Superhard and Novel Carbon Materials, 7a Centralnaya str., Troitsk, Moscow region, 142190, Russia*
*3) L.V. Kirensky Institute of Physics, Akademgorodok, Krasnoyarsk, 660036, Russia*

e-mail: sergbsi@gmail.com; sburkov@sfu-kras.ru

Anisotropy of dc electric field influence on the different types of acoustic waves in piezoelectric plate has been investigated by means of computer simulation. Detail calculations have made for bismuth germanium oxide crystals.

## INTRODUCTION

Investigation of acoustic wave propagation in the piezoelectric plates under the bias electric field leads to the possibility of the controlling of acoustoelectonic devices parameters. Detail theory of the dc electric field or mechanical stress influences on the bulk acoustic (BAW) and surface acoustic (SAW) waves propagation in piezoelectric crystals had been derived in [1-6]. The studying of the anisotropic propagation of the zero order Lamb wave in the lithium niobate piezoelectric plates has fulfilled by the authors [7, 8]. Peculiarities of Lamb waves and surface waves with the horizontal polarization (SH) propagating along the high symmetry directions of the cubic piezoelectric crystals under the dc electric field have been investigated earlier [9].

To optimize the acoustoelectronic device it is necessary to find both an appropriate crystal direction and a value of the (h×f) product for a given frequency range (h – the crystalline plate thickness, f - the frequency). In the present paper the anisotropy of Lamb and SH-waves parameters in the bismuth germanium oxide ($Bi_{12}GeO_{20}$) piezoelectric crystal under the influence of dc electric field has been investigated by means of computer simulation.

## PROPAGATION THEORY OF LAMB AND SH- WAVES IN PIEZOELECTRIC PLATE UNDER THE INFLUENCE OF HOMOGENEOUS DC ELECTRIC FIELD

Influence of homogeneous dc electric field E on Lamb and SH wave propagation conditions in piezoelectric crystalline plate has been considered on the basis of the theory of bulk acoustic waves propagation in piezoelectric crystals subjected to the action of a bias electric field [1].

Wave equations and electrostatics equation written in the natural state for homogeneously deformed crystals without center of symmetry have the form [2]:

$$\rho_0 \ddot{\tilde{U}}_i = \tilde{\tau}_{ik,k},$$
$$\tilde{D}_{m,m} = 0. \tag{1}$$

Here $\rho_0$ is the density of crystal taken in non-deformed (initial) state, $\tilde{U}_i$ is the vector of dynamic elastic displacements, $\tau_{ik}$ is the tensor of thermo-dynamical stresses and $D_m$ is the vector of the induction of electricity. Here and further the tilde sign is marked the time dependent variables. Comma after the lower index denotes a spatial derivative and Latin coordinate indexes are changed from 1 to 3. Here and further the summation on twice recurring lower index is understood.

State equations can be written as:
$$\tilde{\tau}_{ik} = C^*_{ikpq}\tilde{\eta}_{pq} - e^*_{nik}\tilde{E}_n,$$
$$\tilde{D}_n = e^*_{nik}\tilde{\eta}_{ik} + \varepsilon^*_{nm}\tilde{E}_m, \tag{2}$$

where $\eta_{AB}$ is the deformation tensor and effective elastic, piezoelectric, dielectric constants are defined by:
$$C^*_{iklm} = C^E_{iklm} + (C^E_{iklmpq}d_{jpq} - e_{jiklm})M_j E,$$
$$e^*_{nik} = e_{nik} + (e_{nikpq}d_{jpq} + H_{njpq})M_j E, \tag{3}$$
$$\varepsilon^*_{nm} = \varepsilon^\eta_{nm} + (H_{nmik}d_{jik} + \varepsilon^\eta_{nmj})M_j E.$$

In (3) E is the value of dc electric field applied to the crystal, $M_j$ is the unit vector of E-direction, $C^E_{iklmpq}$, $e_{nikpq}$, $\varepsilon^\eta_{nmj}$, $H_{nmik}$ are nonlinear elastic, piezoelectric, dielectric and electrostrictive constants (material tensors), $d_{jpq}$ and $e_{nik}$ are the piezoelectric tensors, $C^E_{iklm}$ and $\varepsilon^\eta_{nm}$ are elastic and dielectric tensors. Substituting (2) into (1) we can obtain Green-Christoffel equation in a general form which can be used for the analysis of bulk acoustic waves propagation in the case of E-influence.

Let's use coordinate system $X_3$ axis directs along the external normal to the surface of a media occupying the space $h \geq X_3 \geq 0$, and the wave propagation direction coincides with $X_1$ axis. Plane waves propagating in the piezoelectric plate are taken in the form:
$$\tilde{U}_i = \alpha_i \exp[i(k_j x_j - \omega t)],$$
$$\varphi = \alpha_4 \exp[i(k_j x_j - \omega t)], \tag{4}$$

where $\alpha_i$ and $\alpha_4$ are amplitudes of elastic wave and electric potential $\varphi$ concerned closely with the wave, and $k_j$ are components of wave propagation vector. Taking into account (2) and (3) the substitution (4) into (1) gives us equation in a specific form. So if the electric field is applied to piezoelectric crystal, Green-Christoffel equation can be written as
$$[\Gamma_{pq}(E) - \rho_0 \omega^2 \delta_{pq}]\tilde{U}_q = 0, \tag{5}$$

where Green-Christoffel tensor has the form:

$$\begin{aligned}
\Gamma_{pq} &= (C^*_{ipqm} + 2d_{jkq}C^E_{ipkm}M_jE)k_ik_m, \\
\Gamma_{q4} &= e^*_{imq}k_ik_m, \\
\Gamma_{4q} &= \Gamma_{q4} + 2e_{ikm}d_{jkq}M_jk_ik_mE, \\
\Gamma_{44} &= -\varepsilon^*_{nm}k_nk_m.
\end{aligned} \qquad (6)$$

Propagation of acoustic waves in the piezoelectric plate under the E-influence should satisfy to boundary conditions of zero normal components of the stress tensor on the boundaries "crystal-vacuum". Continuity of the electric field components which are tangent to the boundary surface is guaranteed by the condition of the continuity of the electrical potential and normal components of the electric displacement vector:

$$\begin{aligned}
\tau_{3k} &= 0, \quad x_3 = 0; \; x_3 = h; \\
\varphi &= \varphi^{[I]}, \quad x_3 < 0; \\
\varphi &= \varphi^{[II]}, \quad x_3 > h; \\
D &= D^{[I]}, \quad x_3 < 0; \\
D &= D^{[II]}, \quad x_3 > h.
\end{aligned} \qquad (7)$$

Here the upper index «I» is concerned to the half-space $X_3 > h$ and index «II» – to the half-space $X_3 < 0$. Substituting the solutions (4) into equations (7) and neglecting of the terms which are proportional $E^2$ (and higher order ones), finally we have obtained the system of equations useful to analyze the change of the wave's structure arising as a consequence of crystal symmetry variation and new effective constants appearance:

$$\begin{aligned}
&\sum_{n=1}^{8} C_n \left( C^*_{3jkl} k_l^{(n)} \alpha_k^{(n)} + e^*_{k3j} k_k^{(n)} \alpha_4^{(n)} \right) \exp\left(ik_3^{(n)}h\right) = 0; \\
&\sum_{n=1}^{8} C_n \left[ e^*_{3kl} k_l^{(n)} \alpha_k^{(n)} - (\varepsilon^*_{3k} k_k^{(n)} - i\varepsilon_0) \alpha_4^{(n)} \right] \exp\left(ik_3^{(n)}h\right) = 0; \\
&\sum_{n=1}^{8} C_n \left( C^*_{3jkl} k_l^{(n)} \alpha_k^{(n)} + e^*_{k3j} k_k^{(n)} \alpha_4^{(n)} \right) = 0; \\
&\sum_{n=1}^{8} C_n \left[ e^*_{3kl} k_l^{(n)} \alpha_k^{(n)} - (\varepsilon^*_{3k} k_k^{(n)} + i\varepsilon_0) \alpha_4^{(n)} \right] = 0.
\end{aligned} \qquad (8)$$

Here the index n = 1,…4 corresponds to the number of one of the partial waves (4) and $C_n$ are the amplitude coefficients of partial waves.

It can remember that the equations (8) were obtained at the assumption of homogeneity of applied dc electric field without taking into account the edge effects. But these equations take into account all changes of the crystal density and the form of crystal sample arising as a consequence of finite deformation of piezoelectric media under the action of strong dc electric field [2].

# ANISOTROPY OF DC ELECTRIC FIELD INFLUENCE ON THE ACOUSTIC WAVE PARAMETERS IN THE BISMUTH GERMANIUM OXIDE PIEZOELECTRIC PLATE

As a model media the bismuth germanium oxide (23 point symmetry) has been used. Taken into account equations (8) and linear and nonlinear material properties from [2], the computer calculation of the main parameters such as phase velocity, electromechanical coupling coefficient (EMCC), and controlling coefficient of phase velocity

$$\alpha_{v_i} = \frac{1}{v_i(0)}\left(\frac{\Delta v_i}{\Delta E}\right)_{\Delta E \to 0} \qquad (9)$$

has been carried out.

Analyses of the variation of waves parameters was fulfilled for (001) and (110) crystalline planes when dc electric field was directed along some kind of $X_1$, $X_2$ and $X_3$ axes.

### Acoustic wave propagation in the (001) crystalline plane

If acoustic wave is propagating in the (001) crystal plane, the application of dc electric field along $X_1$ axis, i.e. along the acoustic wave propagation direction, or along $X_2$ axis, i.e. orthogonal to the sagittal plane, leads to decreasing of the crystal symmetry to triclinic one in general case except the field directions coinciding with basic axes of the crystal. Last variants have been considered in details earlier [9, 10]. Under the action of dc electric field along $X_3$ basic axis (E||[001]), i.e. orthogonal to the free surface of the crystal plate, the crystal symmetry decreases to monoclinic one (2 point symmetry). Twofold axis coincides with the [001] direction and there are induced some effective elastic constants:

$$\begin{aligned}
C_{16}^* &= (C_{166}d_{14} - e_{124})E; & C_{26}^* &= (C_{155}d_{14} - e_{134})E; \\
C_{36}^* &= (C_{144}d_{14} - e_{114})E; & C_{45}^* &= (C_{456}d_{14} - e_{156})E; \\
e_{15}^* &= (e_{156}d_{14} + H_{55})E; & e_{31}^* &= (e_{124}d_{14} + H_{31})E; \\
e_{32}^* &= (e_{134}d_{14} + H_{32})E; & e_{33}^* &= (e_{114}d_{14} + H_{33})E.
\end{aligned} \qquad (10)$$

Phase velocities of the zero and first order modes of Lamb wave propagating in the (001) crystalline plane with (h×f) values up 500 to 3000 m/s are shown on Fig. 1. If (h×f) values are increased, the phase velocity of antisymmetric mode $A_0$ is considerably increased, for example up 1284.96 до 1645.58 m/s in the [110] direction, but the square of EMCC (Fig. 2) is decreased up 2.3% to 0.8%. Values of the EMCC square were calculated as in the case of SAW propagation, i.e. when the metallization of one free surface of piezoelectric plate has taken into account [11]. For $A_1$ mode propagating along the [110] direction of the (001) plane there is the EMCC square's maximal value equal to 0.5% (h×f = 1500 m/s). Note that the EMCC square's qualitative behavior for the first order modes is similar to the zero order ones, but its numerical values are considerably less.

If E∥$X_1$, dependences for $α_v$ coefficients of $A_0$ mode are similar to ones for Rayleigh surface acoustic wave [12]. When the $A_0$ mode is propagating along the [100] direction, its $α_v$ coefficients are considerably increased up $-4.17·10^{-11}$ to $-2.5·10^{-10}$ m/V by the variation of (h×f) quantity up 500 to 3000 m/s. For the $A_1$ mode maximal $α_v$ values are reached in the [110] propagation direction, in particular $α_v = -6.7·10^{-11}$ m/V (h×f = 2500 m/s).

Note that as a result of E application orthogonal to the sagittal plane (E∥$X_2$), the [100] and [010] propagation directions, undistinguished for the undisturbed crystal, become unequal ones. This effect is the consequence of the 23 point symmetry peculiarity of the given crystal, since there is a difference between components of nonlinear properties which are responsible for the E influence on the phase velocities, for example $C_{155} ≠ C_{166}$, $e_{124} ≠ e_{134}$, $H_{12} ≠ H_{21}$. It can point to the fact that these components are equal in all other piezoelectric crystals of cubic symmetry. So $α_v = -6.1·10^{-12}$ m/V and $α_v = -7.3·10^{-12}$ m/V for the $A_0$ modes in the [100] and [010] directions (Fig. 2, c). In the case when both main surfaces of the plate are coated by metal, the E application along $X_3$ axis leads to increasing of $α_v$ coefficients as a result of the thickness increasing. In particular $A_0$ mode ([110] propagation direction) $α_v$ coefficient varies up $-2.58·10^{-10}$ to $-3.71·10^{-10}$ m/V (Fig. 2).

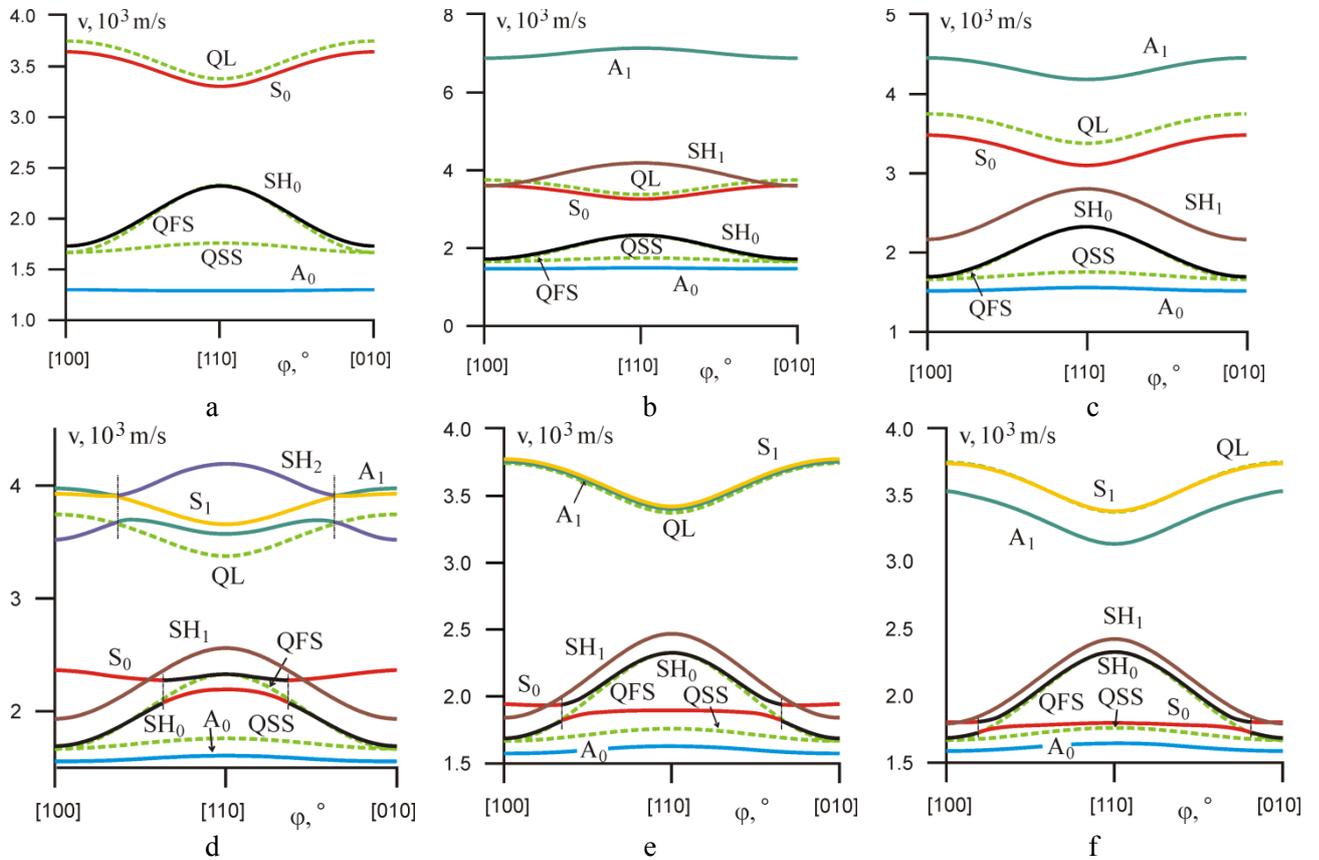

Fig. 1. Phase velocities of acoustic waves propagating in the (001) plane of $Bi_{12}GeO_{20}$ crystal at E = 0 and various values of h×f (m/s): a) h×f = 500; b) h×f = 1000; c) h×f = 1500; d) h×f = 2000; e) h×f = 2500; f) h×f = 3000. Curves for the quasi-longitudinal, fast and slow quasi-shear bulk acoustic waves are marked as QL, QFS, QSS respectively.

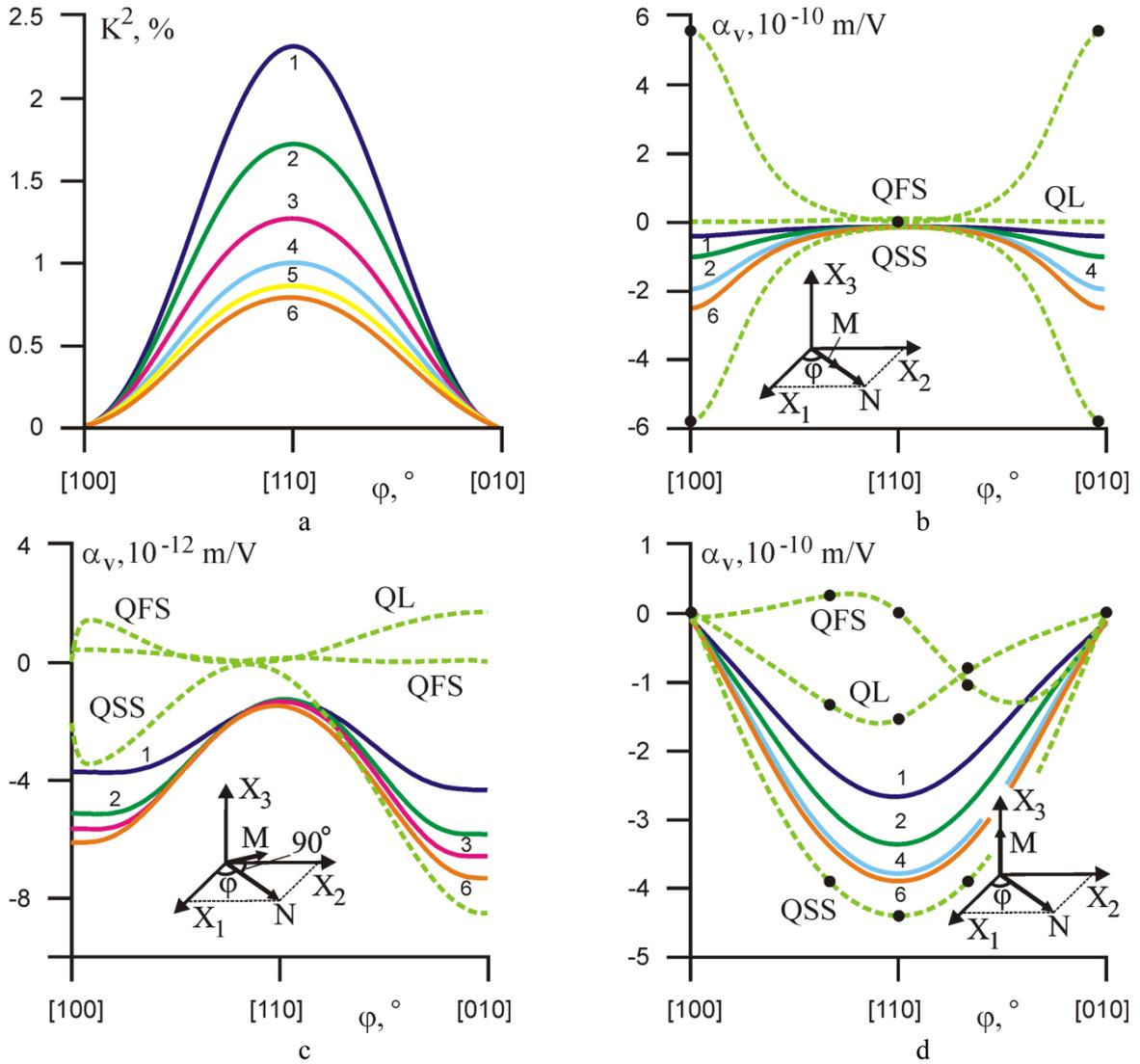

Fig. 2. The square of EMCC and $\alpha_v$ coefficients of bulk waves and the $A_0$ modes propagating in the (001) plane of $Bi_{12}GeO_{20}$ crystal: a) the square of EMCC; b) $\alpha_v$ coefficients ($E\|X_1$); c) $\alpha_v$ coefficients ($E\|X_2$); d) $\alpha_v$ coefficients ($E\|X_3$). A number of the curve corresponds to the values $h\times f$ (m/s): 1 – $h\times f$ = 500; 2 – $h\times f$ = 1000; 3 – $h\times f$ = 1500; 4 – $h\times f$ = 2000; 5 – $h\times f$ = 2500; 6 – $h\times f$ = 3000. Points are marked the experimental $\alpha_v$ coefficients [2].

One of the distinctive peculiarities of wave's propagation in the (001) crystalline plane is the hybridization effect. There are some coupled modes having the energy exchange. The quantitative measure of this effect is the hybridization coefficient [13]:

$$M = \frac{W^{12} + W^{21}}{W^1 + W^2}, \qquad (11)$$

where $W^{12} + W^{21}$ – a complete mutual energy of two coupled modes (time average); $W^1 + W^2$ – a complete energy of acoustic wave. Without electric field the hybridization effect between the $S_0$ и $SH_0$ modes exists only in the thick plates if $h\times f \geq 2000$ m/s (Fig. 3). It should be noted that the hybridization takes place in the point of the equality of phase velocities of the $S_0$ and $SH_0$ modes with the phase velocity of the QFS bulk acoustic wave. Hybridization regions are shown by vertical lines on insets of Fig. 3, b. The E-application along the $X_1$ or $X_2$ axes amplifies the hybridization effect, and $\alpha_v$ values are increased in accordance with exponential function and reach the maximal quanti-ties (Fig. 4, 5). If the electric field is applied along the $X_1$ or $X_2$ axes the hybridization effect arises between the $S_0$ and $SH_1$ modes lacking in the undisturbed case (Fig. 3). The E-application along the $X_3$ axis leads to the hybridization effect decreasing. Maximal $α_v$ values for the $S_0$ mode are reached in the [110] propagation direction of the (001) plane when $E||X_1$ increasing up -1.9·$10^{-12}$ to 4.34·$10^{-10}$ m/V by the increment of (h×f) value (Fig. 4).

If $E||X_3$ maximal $α_v$ values take place in the [110] direction: $α_v$ = -2.99·$10^{-10}$ m/V (h×f = 3000 m/s). Square of EMCC of the $S_0$ mode is considerably increased in the [110] direction if (h×f) values are increased: up 0.47 % to 0.84 % when h×f = 500 и 3000 m/s respectively. Hybridization effect leads to exponential dependence of the $α_v$ coefficient and changes the EMC coefficients in the hybridization region. When $E||X_1$ anisotropy of $α_v$ coefficient for the $SH_0$ mode is the similar one to the $S_0$ mode (Fig. 5). When dc electric field is applied along the $X_3$ or $X_2$ axes the $α_v$ coefficient values don't depend on (h×f) values excluding the hybridization region between the $S_0$ и $SH_0$ modes. Maximal EMCC values (2.8 %) take place in the [100] and [010] directions of the (001) crystalline plane.

The $α_v$ coefficients of the $S_1$, $A_1$, and $SH_1$ first modes propagating in the (001) plane are shown on the Fig. 6. Distinctive peculiarity for higher order modes is the extreme behavior of $α_v$ coefficient in the region when (h×f) value is close to critical one, and the appearance of acoustic modes of higher order becomes possible. In this case a small variation in the plate configuration and material properties of the crystal changes considerably the phase velocity. In particular for the $A_1$ mode propagating along the [110] direction there is maximal value $α_v$ = -12.32·$10^{-10}$ m/V ($E||X_3$). Note that there is the influence of the hybridization effect which leads to the exponential dependence of the $α_v$ coefficient.

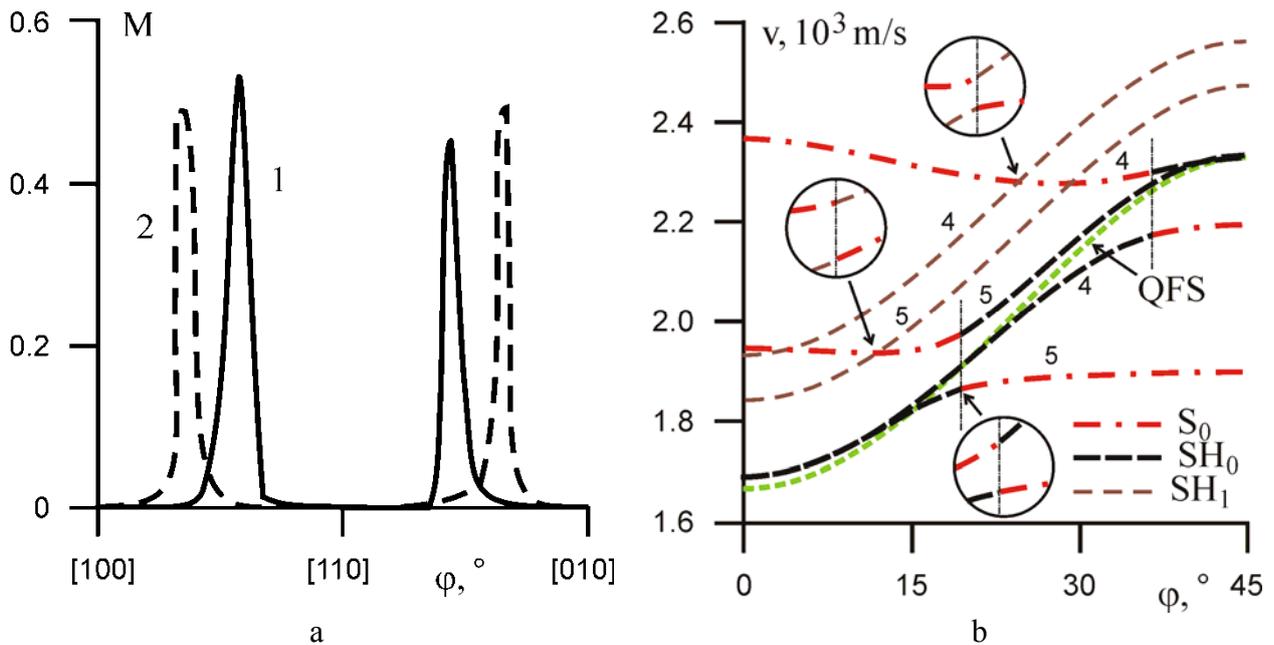

Fig. 3. a) Hybridization coefficient M ($E||X_1$) for: 1 - $S_0$-$SH_1$ modes (h×f = 2000 m/s); 2 – $S_0$-$SH_0$ (h×f = 2500 m/s). b) Phase velocities of acoustic waves. Designations of curves are in accordance with Fig. 2.

## Acoustic wave propagation in the (110) crystalline plane

For this plane the anisotropy of phase velocities is shown on Fig. 7 and the maximal values of the $\alpha_v$ coefficients are realized if $E \| X_1$. The E application along the $X_3$ axis has a minimal effect on the phase velocities of the $A_0$, $S_0$ and $SH_0$ modes (Fig. 8, 9). Maximal value $\alpha_v = -5.8 \cdot 10^{-10}$ m/V for $S_0$ mode takes place in the direction oriented relative to the [001] axis under the angle $\varphi = 29°$ ($h \times f = 1000$ m/s) (Fig. 9). Distinctive features of the $\alpha_v$ coefficients in the (110) plane for the bulk acoustic waves are defined by the splitting of the tangent type acoustic axis coinciding with the [001] crystalline direction (twofold axis of symmetry) in the undisturbed state and by the displacement of the conic type acoustic axis coinciding with the [111] crystalline direction (three-hold axis of symmetry) in the undisturbed state [2].

Extreme values of the $\alpha_v$ coefficients are shown in the Table.

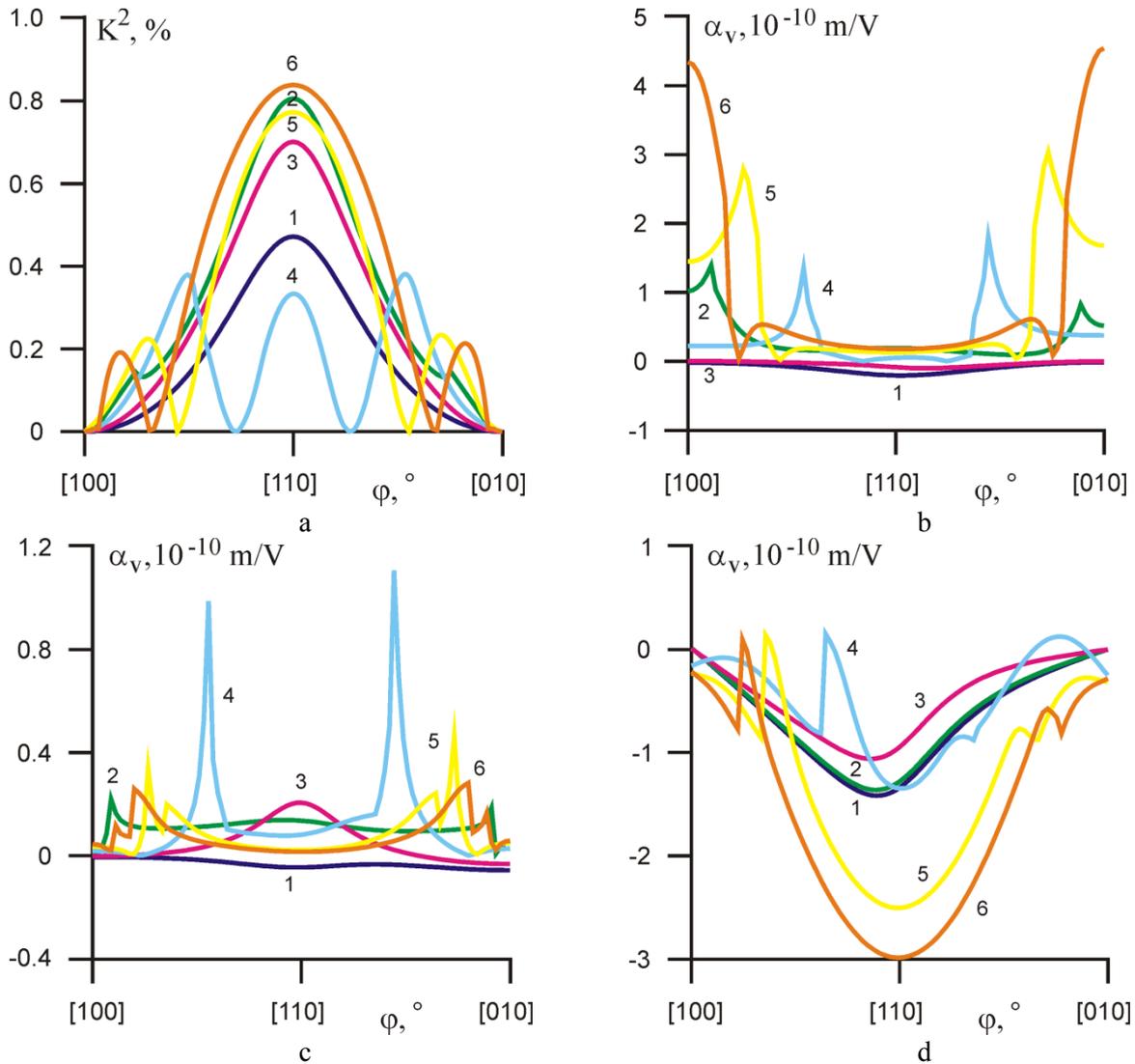

Fig. 4. Square of EMCC (a) and $\alpha_v$ coefficients of the $S_0$ mode propagating in the (001) plane of $Bi_{12}GeO_{20}$ crystal: b) – $E \| X_1$; c) – $E \| X_2$; d) – $E \| X_3$. A number of the curve corresponds to the values $h \times f$ (m/s): 1 – $h \times f = 500$; 2 – $h \times f = 1000$; 3 – $h \times f = 1500$; 4 – $h \times f = 2000$; 5 – $h \times f = 2500$; 6 – $h \times f = 3000$.

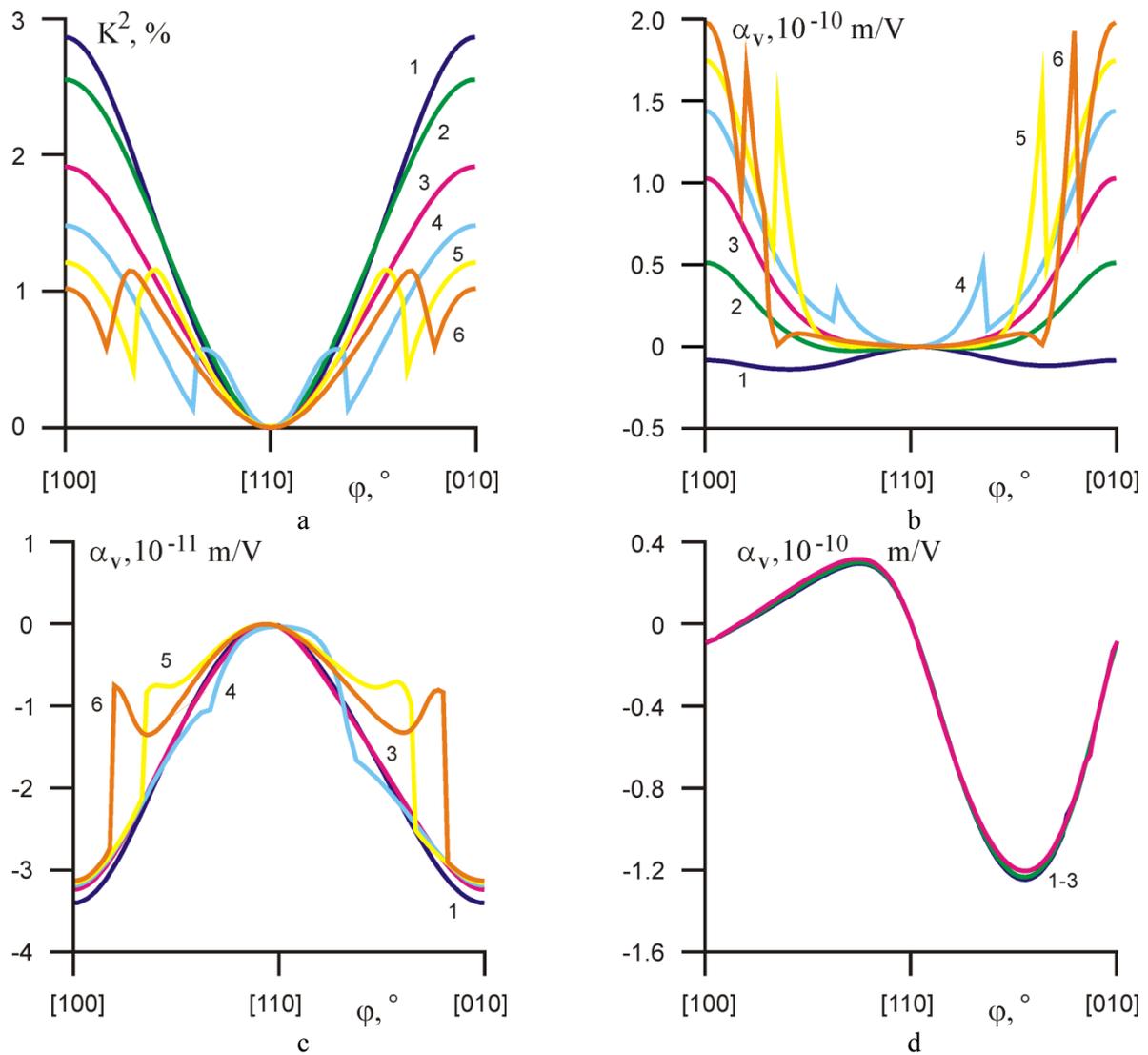

Fig. 5. Square of EMCC (a) and $\alpha_v$ coefficients of the $SH_0$ mode propagating in the (001) plane of $Bi_{12}GeO_{20}$ crystal: b) – $E\|X_1$; c) – $E\|X_2$; d) – $E\|X_3$. A number of the curve corresponds to the values $h\times f$ (m/s): 1 – $h\times f = 500$; 2 – $h\times f = 1000$; 3 – $h\times f = 1500$; 4 – $h\times f = 2000$; 5 – $h\times f = 2500$; 6 – $h\times f = 3000$.

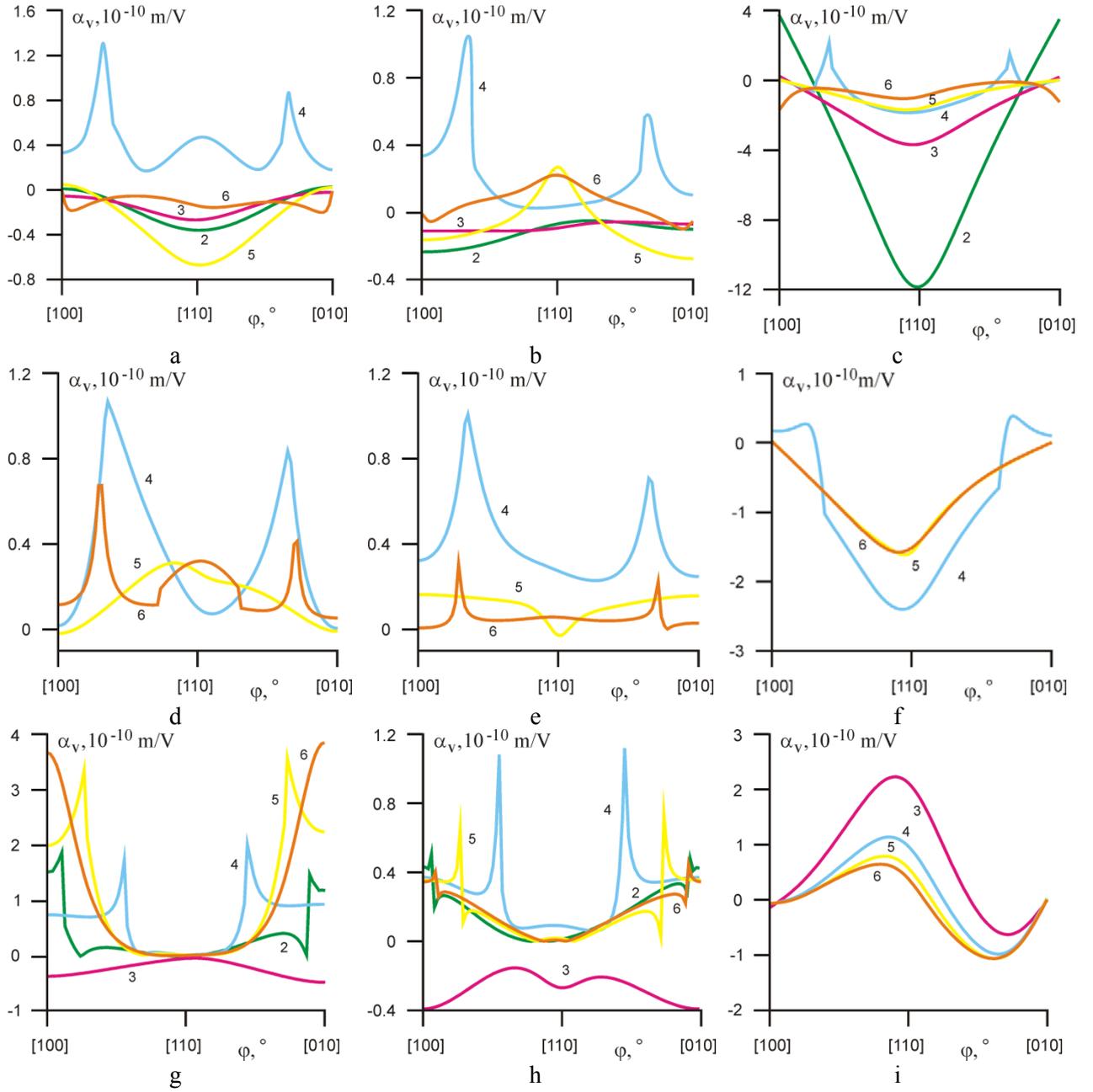

Fig. 6. The $\alpha_v$ coefficients of the first Lamb and SH modes propagating in the (001) plane of $Bi_{12}GeO_{20}$ crystal: (a – c) - $A_1$; (d – f) - $S_1$; (g – i) - $SH_1$; a, d, g – $E\|X_1$; b, e, h – $E\|X_2$; c, f, i – $E\|X_3$. Designations of curves are in accordance with Fig. 2.

Table. Maximal and minimal values of the $\alpha_v$ coefficients of Lamb and SH waves in the bismuth germanium oxide crystal

| Crystalline plane | Mode | DC electric field direction | Angle $\varphi$ | h×f, m/s | $\alpha_v$, $10^{-11}$ m/V |
|---|---|---|---|---|---|
| (001) | $A_0$ | E‖$X_2$ | 47 | 500 | -0.132 |
| | | E‖$X_3$ | 45 | 3000 | -39 |
| | $SH_0$ | E‖$X_1$ | 0 | 3000 | 19.9 |
| | | E‖$X_3$ | 70 | 500 | -12.5 |
| | $S_0$ | E‖$X_1$ | 90 | 3000 | 45.4 |
| | | E‖$X_3$ | 45 | 3000 | -29.9 |
| | $A_1$ | E‖$X_3$ | 0 | 1000 | 36.2 |
| | | E‖$X_2$ | 90 | 2500 | -2.74 |
| | $SH_1$ | E‖$X_1$ | 90 | 3000 | 38.6 |
| | | E‖$X_3$ | 72 | 3000 | -10.6 |
| | $S_1$ | E‖$X_1$ | 16 | 2000 | 10.7 |
| | | E‖$X_3$ | 42 | 2000 | -24 |
| (110) | $A_0$ | E‖$X_1$ | 90 | 1000 | 30.1 |
| | | E‖$X_1$ | 30 | 500 | -17.9 |
| | $SH_0$ | E‖$X_2$ | 67 | 1000 | 19.6 |
| | | E‖$X_1$ | 90 | 1000 | -50.3 |
| | $S_0$ | E‖$X_2$ | 51 | 1000 | 9.69 |
| | | E‖$X_1$ | 29 | 1000 | -58.1 |

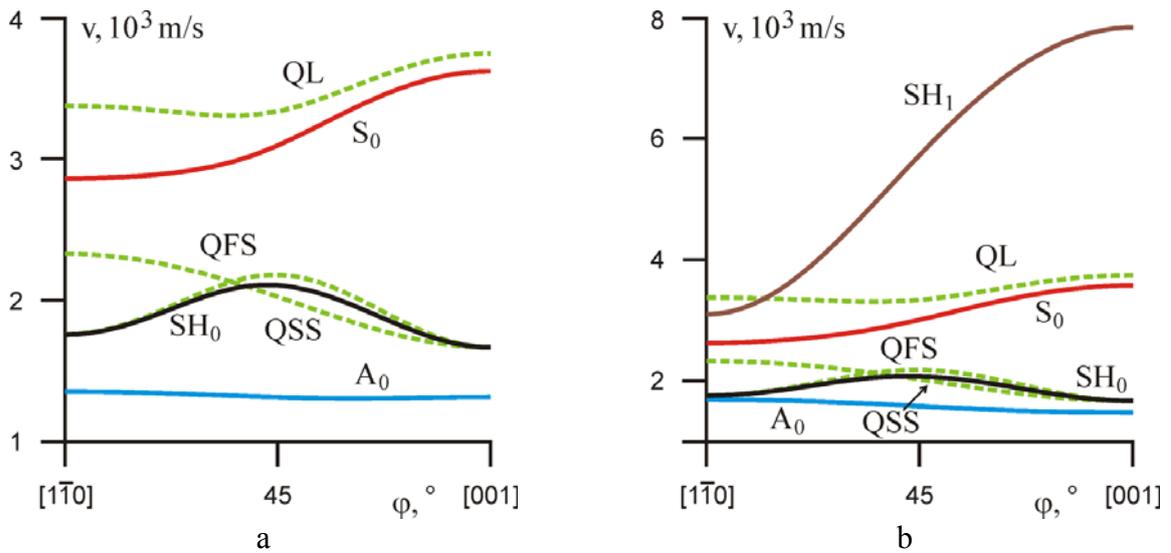

Fig. 7. Phase velocities of acoustic waves propagating in the (110) plane of $Bi_{12}GeO_{20}$ crystal at E = 0 and various values of h×f (m/s): a) h×f = 500; b) h×f = 1000. Curves for the quasi-longitudinal, fast and slow quasi-shear bulk acoustic waves are marked as QL, QFS, QSS respectively.

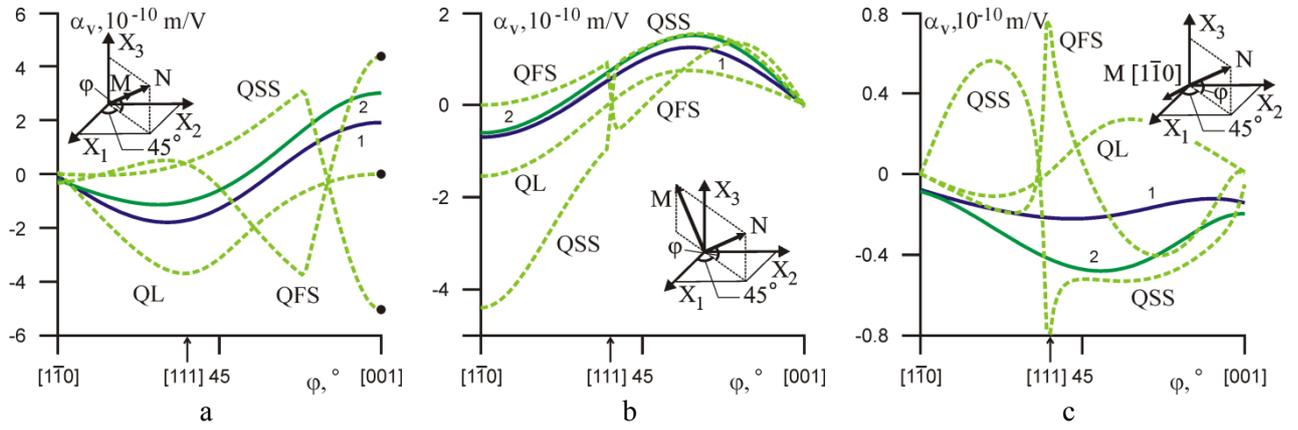

Fig. 8. The $\alpha_v$ coefficients of the $A_0$ mode propagating in the (110) plane of $Bi_{12}GeO_{20}$ crystal: a) $E\|X_1$; b) $E\|X_2$; c) $E\|X_3$. Designations of curves are in accordance with Fig. 2.

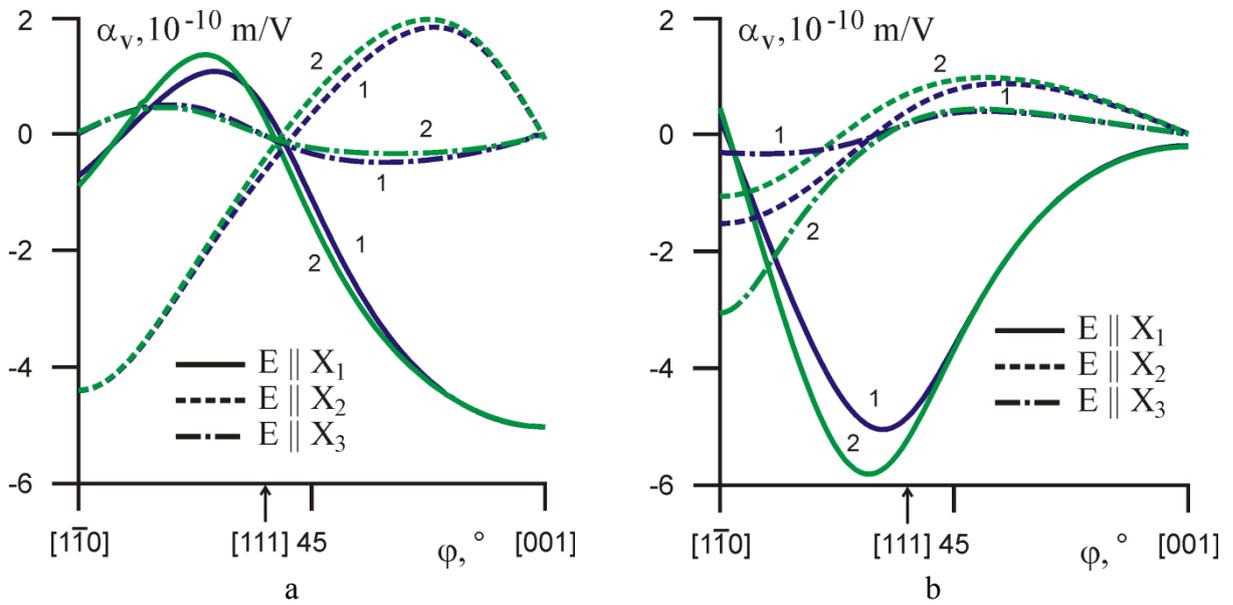

Fig. 9. The $\alpha_v$ coefficients of the $SH_0$ (a) and $S_0$ (b) modes propagating in the (110) plane of $Bi_{12}GeO_{20}$ crystal. Designations of curves are in accordance with Fig. 2.

CONCLUSION

Thus, the anisotropy of homogeneous dc electric field influence on the different types of acoustic waves in the bismuth germanium oxide piezoelectric crystal plate has been investigated by means of computer simulation. Detail analysis of the dispersive behavior of zero and first order Lamb and SH modes has been carried out. Crystalline directions with extreme dc electric field influence have found. It was shown that the acoustic modes interaction can arise as a consequence of dc electric field action in the some directions. The obtained data can be useful to design the controlling devices of acoustoelectronics.

This paper was supported by the Russia President's Program on leading scientific schools support (grant # 4645.2010.2)

# LITERATURE

1. Zaitseva M.P., Kokorin Yu.I., Sandler Yu.M., Zrazhevsky V.M., Sorokin B.P., and Sysoev A.M. *Non-linear Electromechanical Properties of Acentric Crystals*, Novosibirsk: Nauka, Siberian Branch, 1986, 177 p.
2. Aleksandrov K.S., Sorokin B.P., and Burkov S.I. *Effective Piezoelectric Crystals for Acoustoelectronics, Piezotechnics and Sensors*, vol. 2, Novosibirsk: SB RAS Publishing House, 2008, 429 p.
3. Burkov S.I., Sorokin B.P., Glushkov D.A., and Aleksandrov K.S. Theory and computer simulation of the reflection and refraction of bulk acoustic waves in piezoelectrics under the action of an external electric field, *Crystallography Reports*, vol. 50, pp. 986-993, Dec. 2005.
4. Sorokin B.P., Zaitseva M.P., Kokorin Yu.I., Burkov S.I., Sobolev B.V., and Chetvergov N.A. Anisotropy of the acoustic waves velocity in sillenite-structure piezoelectric crystals controlling by the electric field, *Sov. Phys. Acoustics*, vol. 32, pp. 412-415, Sept.-Oct. 1986.
5. Lyamov V.E. *Polarizing Effects and the Anisotropy of Acoustic Waves Interaction in Crystals*, Moscow: MSU Publishing House, 1983, 224 p.
6. Kororin Yu.I., Sorokin B.P., Burkov S.I., and Aleksandrov K.S. Changes of acoustic properties of a cubic piezoelectric crystal in the presence of constant electric field, *Sov. Kristallographiya*, vol. 31, pp. 706-709, Ju.-Aug. 1986.
7. Kuznetsova I.E., Zaitsev B.D., Borodina I.A., et al. Investigation of Acoustic Waves of Higher Order Propagating in Plates of Lithium Niobate, *Ultrasonics*, vol. 42, pp. 179–182. 2004.
8. Kuznetsova I.E., Zaitsev B.D., Joshi S.G., and Borodina I.A. Investigation of Acoustic Waves in Thin Plates of Lithium Niobate and Lithium Tantalate, *IEEE Transactions on Ultrasonics, Ferroelectrics, and Frequency Control*, vol. 48, pp. 322-328. Jan. 2001.
9. Burkov S.I., Zolotova O.P., Sorokin B.P., and Aleksandrov K.S. Effect of External Electrical Field on Characteristics of a Lamb Wave in a Piezoelectric Plate, *Acoustical Physics*, vol. 56, pp. 644–650. Nov.-Dec. 2010.
10. Zolotova O.P., Burkov S.I., and Sorokin B.P. Propagation of the Lamb and SH-waves in Piezoelectric Cubic Crystal's Plate, *Journal of Siberian Federal University. Mathematics & Physics*, vol. 3, pp. 185-204. Mar.-Apr. 2010.
11. Zaitsev B.D., Kuznetsova I.E., Borodina I.A., and Joshi S.G. Characteristics of Acoustic Plate Waves in Potassium Niobate ($KNbO_3$) Single Crystal, *Ultrasonics*, vol. 39, pp. 51-55. 2001.
12. Aleksandrov K.S., Burkov S.I., and Sorokin B.P. Influence of External Homogeneous Electric Field on Properties of Rayleigh Waves in Piezoelectric Plates, *Physics of the Solid State*. vol. 32, pp. 186-192. Jan. 1990.
13. Kuznetsova I.E., Zaitsev B.D., Teplykh A.A., and Borodina I.A. Hybridization of Acoustic Waves in Piezoelectric Plates, *Acoustical Physics*, vol. 53, pp. 73–79. Jan. 2007.